\documentclass[a4paper,12pt]{article}
\usepackage{amssymb}
\usepackage{latexsym}
\usepackage[dvips]{graphicx}
\usepackage{cite}
\newcommand{\be}{\begin{equation}}
\newcommand{\ee}{\end{equation}}
\newcommand{\bea}{\begin{eqnarray}}
\newcommand{\nn}{\nonumber}
\newcommand{\eea}{\end{eqnarray}}

\begin{document}

\begin{titlepage}
\begin{flushright}
\end{flushright}
\begin{centering}
\vspace{.3in}
{\Large{\bf Gravitational Anomalies, Hawking Radiation,
and  Spherically Symmetric Black Holes}}
\\

\vspace{.5in} {\bf  Elias C.
Vagenas$\;^{a,}$\footnote{Email: evagenas@phys.uoa.gr}
and
Saurya Das$\;^{b,}$\footnote{Email: saurya.das@uleth.ca}
}\\
\vspace{0.3in}

$\;^a$Nuclear and Particle Physics Section\\
Physics Department\\
University of Athens\\
GR-15771, Athens, Greece\\
\vspace{.1in}
$\;^b$Department of Physics\\
University of Lethbridge\\
4401 University Drive, Lethbridge\\
Alberta - T1K 3M4, Canada\\

\end{centering}

\vspace{0.7in}
\begin{abstract}
\par\noindent
Motivated by the recent work of Robinson and Wilczek,
we evaluate the gravitational anomaly of a chiral scalar field
in a Vaidya spacetime of arbitrary mass function,
and thus the outgoing flux from the time-dependent horizon
in that spacetime.
We show that this flux differs from that of
a perfect blackbody at a fixed temperature.
When this flux is taken into account,
general covariance in that spacetime
is restored. We also generalize their
results to the most general static, and spherically symmetric spacetime.

\end{abstract}

\end{titlepage}
\newpage

\baselineskip=18pt
\section*{Introduction}
Classically, the energy-momentum tensor of any field is expected to be
covariantly conserved in a curved background. Quantum mechanically, however,
this is not always the case. For example, for a chiral scalar field
in $(1+1)$-dimensional curved spacetime, the covariant derivative of the
energy-momentum tensor reads
\be
\nabla_{\mu} T^{\mu}_{\nu}=\frac{1}{96\pi\sqrt{-g}}\epsilon^{\beta\delta}
\partial_{\delta}\partial_{\alpha}
\Gamma^{\alpha}_{\nu\beta}
\hspace{1ex},
\ee
the right hand side being the gravitational anomaly in that spacetime
\cite{bert1,bert2,louis}.

Under certain simplifying assumptions,
it was shown by Christensen and Fulling
\cite{Christensen:1977jc}\footnote{For recent applications see
\cite{Christodoulakis:2001ps,Vagenas:2003tv,Solodukhin:2005ah,Setare:2000ba,Setare:2000py}.},
that the above anomaly can be interpreted
as a flux of radiation, which
quantitatively agrees with the Hawking flux
\cite{Hawking:1974sw,Hawking:1974rv},
from a horizon in that spacetime.
This means that Hawking radiation is a necessary consequence of quantization
\cite{Gibbons:1976ue,Peet:2000hn}
(just as anomaly is),
and that it also helps to restore
general covariance. The resultant `total' energy-momentum
tensor is covariantly conserved.

Recently, the above idea was re-visited by Robinson and Wilczek, who
demonstrated that the result was valid for a wide variety
of spacetimes, and without many of the previous assumptions
(henceforth abbreviated as the R-W method)
\cite{Robinson:2005pd} \footnote{For more details see \cite{phd}}.
Thus Hawking radiation indeed restores general covariance for
a large class of spacetimes.
The spacetimes considered in the R-W method encompassed many of the
known spherically symmetric black hole solutions. However, it
excluded certain
others, such as the Garfinkle-Horowitz-Strominger (GHS) black hole
in string theory. Furthermore, black holes with non-static horizons,
such as the Vaidya spacetime, were excluded as well.
In this paper, we show that the R-W method can be applied to
both the above scenarios. For the most general spherically
symmetric black hole (including the GHS black hole), the
outgoing flux that is dictated by gravitational anomaly
agrees with the flux from a perfect blackbody, radiating at the
Hawking temperature of the black hole. For Vaidya spacetime,
although such a flux exists, it does not agree with a perfect
blackbody flux. The reason of course is that the Hawking temperature of
the black hole is no longer constant in time.
Turning the argument around, one can say that the flux that may be
observed from an evolving horizon is the one above.
This paper is organized as follows.
In Section 1 we review the R-W method for the case of Schwarzschild black hole.
In Section 2 we extend the R-W method to the case of nonstatic Vaidya
spacetime of arbitrary time-dependent mass function
and derive its flux.
In Section 3 we generalize the R-W method to the most
static, and spherically symmetric spacetimes.
As an example of the generalized method, we study the
stringy GHS black hole.
It is shown that the gravitational anomaly in
this stringy black hole background is cancelled
by the total flux of a $1+1$ dimensional blackbody
at the Hawking temperature of this stringy black hole.
Finally, Section 4 is devoted to a brief summary of our results.
\section{R-W method for the Schwarzschild type black holes}
Robinson and Wilczek \cite{Robinson:2005pd} considered a
$d$-dimensional Schwarzschild type spacetime with the metric
\be
ds^2 = -f(r)dt^2 +\frac{dr^2}{f(r)} +r^2 d\Omega^2_{(d-2)}
\label{metric1}
\ee
where $f(r)$ is arbitrary and
$d\Omega^2_{(d-2)}$ is the metric on $S^{d-2}$.
This metric describes many interesting solutions of Einstein equations.
We assume that it has a single non-degenerate horizon at
$r=r_H$.

\par\noindent
Now, the classical action functional for gravity coupled to matter,
$S[matter, g_{\mu\nu}]$, under general coordinate transformations,
changes as
\be
\delta_{\lambda}S=-\int d^{d}x\sqrt{-g}\lambda^{\nu}\nabla_{\mu}T^{\mu}_{\nu}
\ee
(where $\lambda$ is the variational parameter),
and that the symmetry of the classical action requires that
\be
\delta_{\lambda}S= 0 \Rightarrow \nabla_{\mu}T^{\mu}_{\nu}=0
\hspace{1ex}.
\ee
R-W propose however, that to avoid problems of divergence associated
with the Boulware vacuum, propagating modes along one lightlike
direction are absent. The price to pay is that the resultant theory
is chiral, for which the above condition is violated
quantum mechanically due to chiral anomaly.
%
Now instead, the general covariance of the full
quantum theory requires the variation of the effective action
$W[g_{\mu\nu}]$ to be zero
\be
\delta_{\lambda}W=0
\hspace{1ex}.
\label{varn0}
\ee
Explicit variation yields
\be
-\delta_{\lambda}W= \int d^2 x \sqrt{-g} \lambda^\nu \nabla_\mu T^\mu_\nu~,
\label{varn1}
\ee
where
%
%
\be
T^{\mu}_{\nu}=T^{\mu}_{i\,\nu}\Theta_{-}
+ T^{\mu}_{o\,\nu}\Theta_{+}
+ T^{\mu}_{\chi\,\nu}H~.
\label{emtensor}
\ee
$\Theta_{\pm}=\Theta\left(\pm (r-r_{H})-\epsilon\right)$
are step functions and
$H \equiv 1- \Theta_{+}-\Theta_{-}$,
which is equal to unity between $r_{H}\pm \epsilon$
and zero elsewhere. $T^{\mu}_{i\,\nu}$ and $T^{\mu}_{o\,\nu}$
are covariantly conserved inside and outside the horizon respectively.
However, $T^{\mu}_{\chi\,\nu}$ in (\ref{emtensor}) is
not conserved due to the chiral anomaly at the horizon,
which is timelike and given by \cite{bert2}
\be
\nabla_{\mu}T_{\chi\;\nu}^{\mu}\equiv A_{\nu}\equiv\frac{1}{\sqrt{-g}}\partial _{\mu}N^{\mu}_{\nu}
\label{anomaly}
\ee
where
\be
N^{\mu}_{\nu}=\frac{1}{96\pi}\epsilon^{\beta\mu}\partial_\alpha \Gamma^{\alpha}_{\nu\beta}
\label{Nquantity}
\ee
and $\epsilon^{\beta\mu}$ is the two dimensional Levi-Civita tensor.
Eq.(\ref{varn1}) can be simplified as
\bea
-\delta_\lambda W
%
&=&\int d^{2}x \sqrt{-g}\lambda^{t}\left\{\partial_{r}\left(N^{r}_{t}H\right)+
\left(T^{r}_{o\,t}-T^{r}_{\chi\,t}+N^{r}_{t}\right)\partial\Theta_{+} +
\left(T^{r}_{i\,t}-T^{r}_{\chi\,t}+N^{r}_{t}\right)\partial\Theta_{-}\right\}\nn\\
&&+\int d^{2}x \sqrt{-g}\lambda^{r}
\left\{\left(T^{r}_{o\,r}-T^{r}_{\chi\,r}\right)\partial\Theta_{+} +
\left(T^{r}_{i\,r}-T^{r}_{\chi\,r}\right)\partial\Theta_{-}\right\}
\hspace{1ex},
\label{vareffaction2}
\eea
which when combined with Eq.(\ref{varn0}) yields the following solution
%
%
%
\bea
T^{t}_{t}&=&-\frac{(K+Q)}{f}-\frac{B(r)}{f}-\frac{I(r)}{f}+T^{\alpha}_{\alpha}(r),\nn\\
T^{r}_{r}&=&\frac{(K+Q)}{f}+\frac{B(r)}{f}+\frac{I(r)}{f},\label{emcpts1}\\
T^{r}_{t}&=&-K+C(r)=-f^{2}T^{t}_{r},\nn
\eea
where 
\bea
B(r)&=&\int^{r}_{r_{H}}f(x)A_{r}(x)dx,\\
C(r)&=&\int^{r}_{r_{H}}A_{t}(x)dx,\\
I(r)&=&\frac{1}{2}\int^{r}_{r_{H}}T^{\alpha}_{\alpha}(x)f'(x)dx
\hspace{1ex},
\eea
and $K,Q$ are constants of integration.
Here, it is assumed that
$\frac{I}{f} \Big|_{r_{H}}=\frac{1}{2}T^{\alpha}_{\alpha}\Big|_{r_{H}}$ is finite,
and
\be
\lim_{(r-r_{H})\rightarrow 0_-}\left(\frac{1}{f}\right)=-\lim_{(r-r_{H})\rightarrow 0_+}\left(\frac{1}{f}\right)
\label{antisym}
\hspace{1ex}.
\ee

%

\par\noindent
Next, in
the limit $\epsilon\rightarrow 0$,
using  Eq.(\ref{antisym}) and
\be
\partial_{\mu}\Theta_{\pm}=\delta^{r}_{\mu}\left(\pm1-\epsilon\partial_{r}
\pm\frac{1}{2}\epsilon^{2}\partial^{2}_{r}-\ldots\right)\delta\left(r-r_{H}\right),
\ee
the variation of the effective action (\ref{vareffaction2}) takes the form
\bea
-\delta_{\lambda}W&=&
\int d^{2}x\lambda^{t}\left\{\left[K_{o}-K_{i}\right]
\delta\left( r-r_{H} \right)\right.
\nn\\
&&-\left. \epsilon\left[K_{o}+K_{i}-2K_{\chi}-2N^{r}_{t}\right]\partial\delta\left( r-r_{H} \right)+\ldots\right\}\nn\\
&&- \int d^{2}x\lambda^{r}\left\{\left[\frac{K_{o}
+ Q_{o}+K_{i}+Q_{i}-2K_{\chi}-2Q_{\chi}}{f}\right]\right.
\delta\left( r-r_{H} \right)
\nn\\
&&-\left.\epsilon\left[\frac{K_{o} +Q_{o}-K_{i}-Q_{i}}{f}\right]\partial\delta\left(r-r_{H}\right)+\ldots\right\}
\label{vareffaction3}
\hspace{1ex}.
\eea
It is easily seen in equation (\ref{vareffaction3}) that the
values of the energy-momentum tensor on the horizon
contribute to the gravitational anomaly.
Also, the parameters $\lambda^{t}$ and $\lambda^{r}$ being independent,
the necessary and sufficient conditions for Eq.(\ref{vareffaction3})
to hold are
\bea
K_{o}&=&K_{i}=K_{\chi}+\Phi\\
Q_{o}&=&Q_{i}=Q_{\chi}-\Phi~,
\label{conditions}
\eea
where
\be
\Phi=N^{r}_{t}\Big|_{r_{H}}
\label{phi}
\hspace{1ex}.
\ee
The energy-momentum tensor now assumes the form
\be
T^{\mu}_{\nu}=T^{\mu}_{c\,\nu}+T^{\mu}_{\Phi\,\nu}~,
\ee
where $T^{\mu}_{c\,\nu}$ represents the conserved energy-momentum tensor
without any quantum effects, and
$T^{\mu}_{\Phi\,\nu}$ is a conserved tensor with $K=-Q=\Phi$, representing the
flux $\Phi$.
\par\noindent
For the specific Schwarzschild type black hole spacetime described by (\ref{metric1}),
one can show that
\bea
N^{t}_{t}&=&N^{r}_{r}=0\nn\\
N^{r}_{t}&=&\frac{1}{192\pi}\left(f'^2 + f''f\right)\\
N^{t}_{r}&=&-\frac{1}{192\pi f^2}\left(f'^2 - f''f\right)\nn
\hspace{1ex},
\eea
implying
\bea
\Phi&=&N^{r}_{t}\Big|_{r_{H}}\\
&=&\frac{1}{192\pi}f'^{2}(r_{H})
\label{flux1}
\hspace{1ex}.
\eea
Now, it
is well known that the surface gravity $\kappa$ in this case is given by
\bea
\kappa&=&\frac{1}{2}\frac{\partial f}{\partial r}\Big|_{r=r_{H}}\\
&=&\frac{1}{2}f'(r_{H})
\label{surface1}
\hspace{1ex},
\eea
which implies the following Hawking temperature
\bea
T_{H}&=&\frac{\kappa}{2\pi}\\
&=&\frac{f'(r_{H})}{4\pi}~.
\eea
On the other hand,
a beam of massless black body radiation moving outwards in the
radial direction at a temperature $T_{H}$ has a flux of the form
\be
\Phi=\frac{\pi}{12}T_H^2
\label{flux2}
\hspace{1ex}.
\ee
Therefore it is evident that the flux (\ref{flux1})
is nothing but the Hawking flux, which exactly cancels the
gravitational anomaly!
%
%
%
Recently, Iso, Umetsu, and Wilczek \cite{Iso:2006wa} showed that in the case of a charged black hole apart from the
gravitational anomalies, gauge anomalies show up. These are cancelled by the Hawking radiation of
charged particles from the charged black hole. Furthermore, extended versions  of
\cite{Robinson:2005pd} were presented in \cite{Iso:2006ut} which included $4$-dimensional rotating black holes as
well as in  \cite{Setare:2006hq} which included the $(2+1)$-dimensional rotating BTZ black hole.
\section{The Vaidya Metric}
In this section we examine
the R-W  method for nonstatic spacetimes, of which one of the
simplest is given by the Vaidya metric
\be
ds^2 =-\left(1-\frac{2M(\upsilon)}{r}\right)d\upsilon^2 +2d\upsilon dr +r^{2}d\Omega^{2}
\label{vmetric}
\ee
where $\upsilon=t+r^\star$ 
is the advanced time coordinate  ($r^\star$ is the tortoise coordinate)
and the mass $M$ is a function of the advanced time $\upsilon$.
This spacetime accommodates two kinds of
surfaces of particular interest.
The apparent horizon is at $r_{AH}=2M$, whereas
the event horizon is denoted by $r_{EH}=r_h$ \cite{Wu:2001nk}.
To determine the null-surface
$r_h=r_h(\upsilon)$, one first defines
\be
\tilde{\upsilon}=\upsilon \hspace{1ex}\mbox{and}\hspace{1ex} \tilde{r}=r-r_h
\label{tortoise}
\hspace{1ex},
\ee
in terms of which the line element (\ref{vmetric})
can be written as \cite{zhong}
\be
ds^2 =-\left(1-\frac{2M(\upsilon)}{r}-2\dot{r}_{h}\right)d\tilde{\upsilon}^2 +2d\tilde{\upsilon}dr +\tilde{r}^{2}d\Omega^{2}~.
\label{vmetric1}
\ee
Then the event horizon $r_h$ satisfies the null-surface condition
\be
1-\frac{2M(\upsilon)}{r_h}-2\dot{r}_h =0~,
\ee
which yields
\be
r_{h}=\frac{2M(\upsilon)}{1-2\dot{r}_h}~,
\hspace{1ex}
\label{vaidyahor1}
\ee
where $\dot{r}_{h}=dr_{h}/d\upsilon$.
The surface gravity is given by
\be
\kappa = \frac{M(\upsilon)}{(1-2\dot{r}_h) r^{2}_{h}}
\ee
and the corresponding radiation temperature by \cite{Li:2000rk}
\be
T=\frac{1-2\dot{r}_{h}}{8\pi M(\upsilon)}~,
\label{vaidtemp}
\ee
which, using (\ref{vaidyahor1}) becomes
\be
T=\frac{1}{4\pi r_h}
\hspace{1ex}.
\label{vaidtemp3}
\ee
It should be noted that since $r_h$ depends on $\upsilon$,
the location of the event horizon as well as the shape of the black hole change with time.
\par\noindent
Since the Vaidya metric can be written in the form of Eq.(\ref{metric1}),
in order to evaluate the corresponding flux one can evaluate
the quantity $N^{r}_{\upsilon}$ on the event horizon.
Using Eq.(\ref{Nquantity}), this is given by
\bea
N^{r}_{\upsilon}&=&\frac{1}{96\pi}\epsilon^{\beta r}\partial_\alpha \Gamma^{\alpha}_{\upsilon\beta}\nn\\
&=&\frac{1}{96\pi}\epsilon^{\upsilon r}\partial_\alpha \Gamma^{\alpha}_{\upsilon\upsilon}\nn\\
&=& \frac{1}{96\pi}\left(\frac{6M^{2}(\upsilon)}{r^{4}}-\frac{2M(\upsilon)}{r^3}\right)
\hspace{1ex}.
\label{quantityN}
\eea
The corresponding gravitational anomaly evaluated on the event horizon
is
\bea
\Phi&=&N^{r}_{\upsilon}\Big|_{r_h}\\
&=&\frac{1}{96\pi}
\left(\frac{6M^{2}(\upsilon)}{r^{4}}-\frac{2M(\upsilon)}{r^3}\right)\Big|_{r_h}\\
&=&\frac{1}{96\pi r_h^2}
\left(\frac{6M^{2}(\upsilon)}{r_{h}^{2}}-\frac{2M(\upsilon)}{r_{h}}\right)
\label{flux3}
\hspace{1ex}.
\eea
\par\noindent
If one now
considers the Vaidya metric to
be radiating at the radiation temperature $T$,
then using Eqs.(\ref{vaidyahor1}) and (\ref{vaidtemp3}), the flux is given by
\be
\Phi=\frac{\pi}{12}T^{2}\left(1-8\dot{r}_{h}+12\dot{r}^{2}_{h}\right)
= \frac{\pi}{12}\xi~T^{2}
\label{flux4}
\hspace{1ex}
\ee
where
\be
\xi \equiv 1-8\dot{r}_{h}+12\dot{r}^{2}_{h}~.
\ee
\par\noindent
Thus, it is seen that the flux from the horizon of
Vaidya spacetime is not the blackbody (thermal) flux
given by (\ref{flux2}). The underlying reason for this
difference is the non-constant temperature in this case, owing
to its time-dependent mass \cite{Singh:2000sp}.
The factor of $\xi$ expresses this dependence, and as expected,
for the special case $M \rightarrow$ a constant,
$\xi \rightarrow 1$ and flux (\ref{flux4}) yields the
previously obtained flux (\ref{flux2}) for a Schwarzschild black hole.

%
\section{Generalizing to non-Schwarzschild type black holes}
In this section, we generalize the R-W method to the
case of non-Schwarzschild type black holes, i.e the
most general static, spherically symmetric
(non-Schwarzschild type black holes) metric
\be
ds^2=-f(r)dt^2 + \frac{dr^2}{g(r)}++r^2 d^{2}\Omega_{(d-2)}
\label{metric2}
\ee
where
\be
f(r)\cdot g(r)\neq 1
\hspace{1ex}.
\ee
If it has a horizon at $r=r_{H}$ then close it, one can write
\bea
f(r)\approx f'(r_{H})\cdot(r-r_{H})\\
g(r)\approx g'(r_{H})\cdot(r-r_{H})
\hspace{1ex}.
\eea
%
%
%
The corresponding surface gravity and the
Hawking temperature are given respectively by
\begin{eqnarray}
\kappa &=& \frac{1}{2}\sqrt{f'(r_{H})g'(r_{H})}
\label{surface2} \\
T_{H} &=& \frac{\sqrt{f'(r_{H})g'(r_{H})}}{4\pi}
\label{temp1}
\hspace{1ex}.
\end{eqnarray}
Thus, a beam of massless blackbody radiation
moving in the positive radial direction at a temperature $T_{H}$
will have a flux of the form
\be
\Phi=\frac{1}{192\pi}f'(r_{H})g'(r_{H})
\label{flux5}
\hspace{1ex}.
\ee
As for R-W, we assume that the physics near the horizon is
described by a $1+1$ dimensional field theory, in the
the `r-t' section of the spacetime (\ref{metric2}),
and as before, the form of the energy-momentum tensor
after variation of the effective action
(\ref{vareffaction2}) is given by
(up to constants $K$, $Q$ and the trace $T^{\alpha}_{\alpha}$)
\bea
T^{t}_{t}&=&-\frac{(K+Q)}{f}-\frac{B(r)}{f}-\frac{I(r)}{f}+T^{\alpha}_{\alpha}(r),\\
T^{r}_{r}&=&\frac{(K+Q)}{f}+\frac{B(r)}{f}+\frac{I(r)}{f},\\
T^{r}_{t}&=&-K+ \bar{C}(r)=-f(r)g(r)T^{t}_{r}.
\label{emcpts2}
\eea
Now the quantities $B(r)$, $C(r)$, and $I(r)$ are defined as follows
\bea
B(r)&=&\int^{r}_{r_{H}}f(x)A_{r}(x)dx,\\
\bar{C}(r)&=&\sqrt{\frac{g(r)}{f(r)}}\int^{r}_{r_{H}}\sqrt{\frac{f(x)}{g(x)}}A_{t}(x)dx,\\
I(r)&=&\frac{1}{2}\int^{r}_{r_{H}}T^{\alpha}_{\alpha}(x)f'(x)dx
\hspace{1ex}.
\eea
Continuing the R-W method, we end up with an identical expression for the
energy-momentum tensor due to the gravitational anomaly,
which is expressed through the pure flux (\ref{phi}).

However the explicit expression associated with the spacetime under consideration is quite different from the one given
for the Schwarzschild type black holes, i.e. expression (\ref{flux1}).
This difference stems from the fact that the components of
$N^{\mu}_{\nu}$ for the
non-Schwarzschild type black holes are now given by
\bea
N^{t}_{t}&=&N^{r}_{r}=0\nn\\
N^{r}_{t}&=&\frac{1}{192\pi}\left(f'g' + f''g\right) \label{genquantN} \\
N^{t}_{r}&=&-\frac{1}{192\pi g^2}\left(g'^2 - g''g\right)\nn
\hspace{1ex}.
\eea
Therefore the quantity $\Phi$ that
describes the pure flux for the non-Schwarzschild type black holes reads
\be
\Phi=\frac{1}{192\pi}f'(r_{H})g'(r_{H})
\label{flux6}
\hspace{1ex}.
\ee
We see that this is identical to the expression (\ref{flux5}),
derived using black hole thermodynamics.
Thus, once again, the gravitational anomaly is cancelled by the
Hawking flux.
\par\noindent
As an application of the above result, we examine the
GHS black hole \cite{ghs}
which is member of a family of solutions to low-energy string theory,
described by the action (in the string frame)
\be
S=\int d^{4}x
\sqrt{-g}\;e^{-2\phi}\left[-R-4\left(\nabla
\phi\right)^{2}+F^2\right]
\ee
where $\phi$ is the dilaton field and $F_{\mu\nu}$ is the Maxwell field associated
with a U(1) subgroup of $E_8 \times E_8$ or ${\it Spin(32)/Z_{2}}$.
Its charged black hole solution is given as
\be
ds^{2}_{\tiny\mbox{string}}=-\frac{\left(1-\frac{2Me^{\phi_0}}{r}\right)}
{\left(1-\frac{Q^{2}e^{3\phi_0}}{M r }\right)}dt
^{2}+\frac{dr^{2}}
{\left(1-\frac{2Me^{\phi_0}}{r}\right)\left(1-\frac{Q^{2}e^{3\phi_0}}
{Mr}\right)}+ r^{2}d\Omega
\label{ghsmetric}
\ee
where $\phi_0$ is the asymptotic constant value of the dilaton.
This metric describes a black hole with an event horizon at
\be
r_{+}=2
M e^{\phi_0}
\ee
when $Q^{2}<2e^{-2\phi_{0}}M^{2}$.
For the aforementioned black hole we have
\bea
f(r)&=&\frac{\left(1-\frac{2Me^{\phi_0}}{r}\right)}
{\left(1-\frac{Q^{2}e^{3\phi_0}}{M r }\right)}
\label{ghselements1}\\
g(r)&=&\left(1-\frac{2Me^{\phi_0}}{r}\right)\left(1-\frac{Q^{2}e^{3\phi_0}}{Mr}\right)
\label{ghselements2}
\hspace{1ex}.
\eea
The corresponding Hawking temperature follows from Eq.(\ref{temp1})
\be
T_{H}=\frac{1}{8\pi M
e^{\phi_{0}}}
\label{ghstemp}
\hspace{1ex}
\ee
when the metric elements
(\ref{ghselements1}) and (\ref{ghselements2}) of GHS black hole are used.
One can see
that the Hawking temperature of the GHS black hole is independent of
the charge $Q$, for $Q<\sqrt{2}e^{-\phi_{0}}M$. \par\noindent At
extremality, i.e. when $Q^{2}=2e^{-2\phi_{0}}M^{2}$, the  GHS black
hole solution (\ref{ghsmetric}) becomes
\be
ds^{2}_{\tiny\mbox{string}}=- dt^{2}+
\left(1-\frac{2Me^{\phi_0}}{r}\right)^{-2}dr^{2}+
r^{2}d\Omega
\hspace{1ex}.
\label{extghsmetric}
\ee
and its Hawking temperature vanishes, since the corresponding
Euclidean section is  smooth without any identifications.
%
The quantity $N^{r}_{t}$ given by (\ref{genquantN}), reads
\be
N^{r}_{t}=\frac{1}{192\pi}\left(g'f'+g f''\right)
\hspace{1ex}.
\ee
When we evaluate this quantity at $r_H$ the second term is zero since $g(r_H)=0$.
Thus,
\bea
N^{r}_{t}\Big|_{r_H}&=&\frac{1}{192\pi}g'(r_H)f'(r_H)\\
&=&\frac{1}{192\pi}\frac{e^{-2\phi}}{4M^2}\\
&=&\frac{\pi}{12}\left(\frac{e^{-2\phi}}{64\pi^{2}M^2}\right)\\
&=&\frac{\pi}{12}\left(\frac{1}{64\pi^{2}M^2 e^{2\phi}}\right)\\
&=&\frac{\pi}{12}\left(\frac{1}{8\pi M e^{\phi}}\right)^2
\hspace{1ex}.
\eea
Therefore, comparing with (\ref{ghstemp}), we get
\be
N^{r}_{t}\Big|_{r_H}=\frac{\pi}{12}T_{H}^{2}
\hspace{1ex}.
\ee
As for the extremal case
(\ref{extghsmetric}), it is obvious that the generalized R-W
method gives the correct null result
(due to the vanishing Hawking temperature, i.e. $T_{H}^{ext}=0$) since in this case $f(r)=1$ and thus $f'(r)=0$.
\section{Conclusions}
In this work, we have
computed the gravitational anomaly for chiral scalar fields for
the nonstatic Vaidya spacetime of arbitrary mass function.
According to R-W, this is the flux of radiation from
a horizon in the above spacetime, such that
general covariance at the quantum level is restored.
To our knowledge, this is the first time
that Hawking flux from such a dynamical spacetime has been computed.
There have been some computations in the past
but only for specific mass functions.
In addition in these cases the flux was in a
rather complicated form, contrary to our result derived here.
In the limiting case where the mass function is equal to the ADM
mass of the Schwarzschild black hole, we recover the R-W results.
Furthermore, we have generalized  their method to
the most general static, and spherically
symmetric spacetimes. We then applied the
generalized method to the
Garfinkle-Horowitz-Strominger stringy black holes.
The gravitational anomaly of this
stringy black hole is cancelled by the flux of a beam of
massless $1+1$ dimensional particles at the Hawking temperature of this
black hole. Moreover,
at extremality we get the known zero temperature and correspondingly
a null flux indicating
that there is no gravitational anomaly.

It would be interesting to study the physical implications of our result for
other dynamical spacetimes, on which we hope to report elsewhere.
\newline

\par\noindent
{\it Note}: A related work by Keiju Murata and Jiro Soda \cite{Murata:2006pt} appeared on the
archive on the same day we submitted our paper.
\section*{Acknowledgements}
The authors would like to thank Frank Wilczek for a first reading of the
paper and for encouragement to publish this work.
Research for ECV is supported by the Greek State Scholarship Foundation (I.K.Y.).
The work of SD was supported in part by the Natural Sciences and Engineering
Research Council of Canada and in part by the Perimeter Institute for Theoretical Physics.
SD would like to thank the Department of Mathematics and Statistics,
University of New Brunswick, where part of the work was done.


\begin{thebibliography}{99}

\bibitem{bert1}
R.~A.~Bertlmann,
{\it Anomalies in Quantum Field Theory}. International
Series of Monographs on Physics {\bf 91}, Clarendon Press, Oxford (1996).

\bibitem{bert2}
R.~A.~Bertlmann and E.~Kohlprath,
Ann.\ Phys. (N.Y.) {\bf 288}, 137 (2001).

\bibitem{louis}
L.~Alvarez-Gaume and E.~Witten,
Nucl.\ Phys.\ B {\bf 234}, 269 (1984).


\bibitem{Christensen:1977jc}
  S.~M.~Christensen and S.~A.~Fulling,
  Phys.\ Rev.\ D {\bf 15} (1977) 2088.

\bibitem{Christodoulakis:2001ps}
  T.~Christodoulakis, G.~A.~Diamandis, B.~C.~Georgalas and E.~C.~Vagenas,
  Phys.\ Rev.\ D {\bf 64}, 124022 (2001)
  [arXiv:hep-th/0107049].

\bibitem{Vagenas:2003tv}
  E.~C.~Vagenas,
  Phys.\ Rev.\ D {\bf 68}, 024015 (2003)
  [arXiv:hep-th/0301149].

\bibitem{Solodukhin:2005ah}
  S.~N.~Solodukhin,
  Phys.\ Rev.\ D {\bf 74}, 024015 (2006)
  [arXiv:hep-th/0509148].

\bibitem{Setare:2000ba}
  M.~R.~Setare and A.~H.~Rezaeian,
  Mod.\ Phys.\ Lett.\ A {\bf 15}, 2159 (2000)
  [arXiv:hep-th/0007121].

\bibitem{Setare:2000py}
  M.~R.~Setare,
  Class.\ Quant.\ Grav.\  {\bf 18}, 2097 (2001)
  [arXiv:hep-th/0011198].

\bibitem{Hawking:1974sw}
  S.~W.~Hawking,
  Commun.\ Math.\ Phys.\  {\bf 43}, 199 (1975)
  [Erratum-ibid.\  {\bf 46}, 206 (1976)].

\bibitem{Hawking:1974rv}
  S.~W.~Hawking,
  Nature {\bf 248} (1974) 30.

\bibitem{Gibbons:1976ue}
  G.~W.~Gibbons and S.~W.~Hawking,
  Phys.\ Rev.\ D {\bf 15}, 2752 (1977).


\bibitem{Peet:2000hn}
  A.~W.~Peet,
  arXiv:hep-th/0008241.


\bibitem{Robinson:2005pd}
  S.~P.~Robinson and F.~Wilczek,
  Phys.\ Rev.\ Lett.\  {\bf 95}, 011303 (2005)
  [arXiv:gr-qc/0502074].

\bibitem{phd}
S.~P.~Robinson, Ph.D. Thesis, M.I.T., 2005\\
(available at http://web.mit.edu/spatrick/www/PhDThesis/ ).


\bibitem{Iso:2006wa}
  S.~Iso, H.~Umetsu and F.~Wilczek,
  Phys.\ Rev.\ Lett.\  {\bf 96}, 151302 (2006)
  [arXiv:hep-th/0602146].

\bibitem{Iso:2006ut}
  S.~Iso, H.~Umetsu and F.~Wilczek,
  Phys.\ Rev.\ D {\bf 74}, 044017 (2006)
  [arXiv:hep-th/0606018].

\bibitem{Setare:2006hq}
  M.~R.~Setare,
   ``Gauge and gravitational anomalies and Hawking radiation of rotating BTZ
  arXiv:hep-th/0608080.

\bibitem{Wu:2001nk}
  S.~Q.~Wu and X.~Cai,
  Int.\ J.\ Theor.\ Phys.\  {\bf 41}, 559 (2002)
  [arXiv:gr-qc/0111045].

\bibitem{zhong}
Li~Zhong-heng, Liang~You and Mi~Li-qin,
Int.\ J.\ Theor.\ Phys.\  {\bf 38}, 925 (1999).

\bibitem{Li:2000rk}
  X.~Li and Z.~Zhao,
  Phys.\ Rev.\ D {\bf 62}, 104001 (2000).

\bibitem{Singh:2000sp}
  T.~P.~Singh and C.~Vaz,
  Phys.\ Lett.\ B {\bf 481}, 74 (2000)
  [arXiv:gr-qc/0002018].

  \bibitem{ghs}
  D.~Garfinkle, G.~T.~Horowitz and A.~Strominger,
  Phys.\ Rev.\ D {\bf 43}, 3140 (1991)
  [Erratum-ibid.\ D {\bf 45}, 3888 (1992)].

\bibitem{Murata:2006pt}
  K.~Murata and J.~Soda,
  arXiv:hep-th/0606069.

\end{thebibliography}
\end{document}